\documentclass{he_symp}
\usepackage{psfig,graphicx,epsfig}
\usepackage{color}
\usepackage{amsmath,amssymb,epic,eepic,array}
\begin{document}
\renewcommand{\FirstPageOfPaper }{ 1}\renewcommand{\LastPageOfPaper }{ 12}

\title{Radio Observations of Supernova Remnants}

\author{W. Reich}

\institute{Max--Planck--Institut f\"ur Radioastronomie, Auf dem H\"ugel 69, 53121 Bonn, Germany}

\maketitle

\begin{abstract}
Supernovae release an enormous amount of energy into the interstellar
medium. Their remnants can observationally be traced up to several
ten-thousand years. So far more than 230 Galactic supernova remnants
(SNRs) have been identified in the radio range. Detailed studies of
the different types of SNRs give insight into the interaction of the
blast wave with the interstellar medium. Shock accelerated particles
are observed, but also neutron stars left from the supernova explosion
make their contribution. X-ray observations in conjunction with radio
data constrain models of supernova evolution.

A brief review of the origin and evolution of SNRs is given, which are
compared with supernova statistics and observational limitations. In
addition the morphology and characteristics of the different types of
SNRs are described, including some recent results and illustrated by
SNRs images mostly obtained with the Effelsberg 100-m telescope.
\end{abstract}

\section{Introduction}

Supernova explosions (SNe) belong to the most spectacular events in
space. Historical SNe could be traced by the naked eye. Systematic
observations of galaxies reveal several events every year, where the SN
is of comparable brightness to the entire galaxy for days up to weeks. SNRs
belong to the strongest radio sources observed. The first
identification of a SNR in the radio range was published by
Bolton et al. (\cite{bol}), associating the Crab nebula with the
radio source Taurus~A. Subsequently many more SNRs have been
identified and studied in great detail at radio wavelength, which were
complemented and extended in recent years by X-ray studies as reviewed
by Bernd Aschenbach (this volume). SNe and SNRs have a major influence
on the properties of the interstellar medium (ISM) and by their
coordinated impact on the evolution of galaxies as a whole. They enrich
the ISM by heavy elements, release about $10^{51}$~ergs and their
blast waves shape and heat the ISM, compress the magnetic field and
accelerate in their shock waves quite efficiently energetic cosmic rays
as observed throughout the Galaxy. They were suggested to trigger star
formation and neutron stars are left from SN events.

\section{Supernova events}
SN have been classified by their optical spectra: SNI with
Balmer lines and SNII without Balmer lines. Meanwhile a number
of sub-groups have been introduced. Important are their explosion
mechanisms depending on the mass of the progenitor stars, as there are
thermo-nuclear burning and core collapse events. A core collapse needs
a more massive star (SNIb/c, SNII-P) than SN from thermo-nuclear burning
(SNIa, SNII-L). Core collapse events leave a neutron star, while from
thermo-nuclear burning a pure blast wave results.

SN rates depend on star formation and therefore the different SN rates depend
on the type of galaxies (Cappellaro et al.
\cite{cap99}). From these results a SN rate in the Milky Way is expected to
be in the range of one SN in about 30 to 50~years. About 80\% of SN
should be core collapse events and only 20\% are expected to release
pure blast waves showing up as radio shells.

\begin{figure*}
\includegraphics[width=17.8cm,clip]{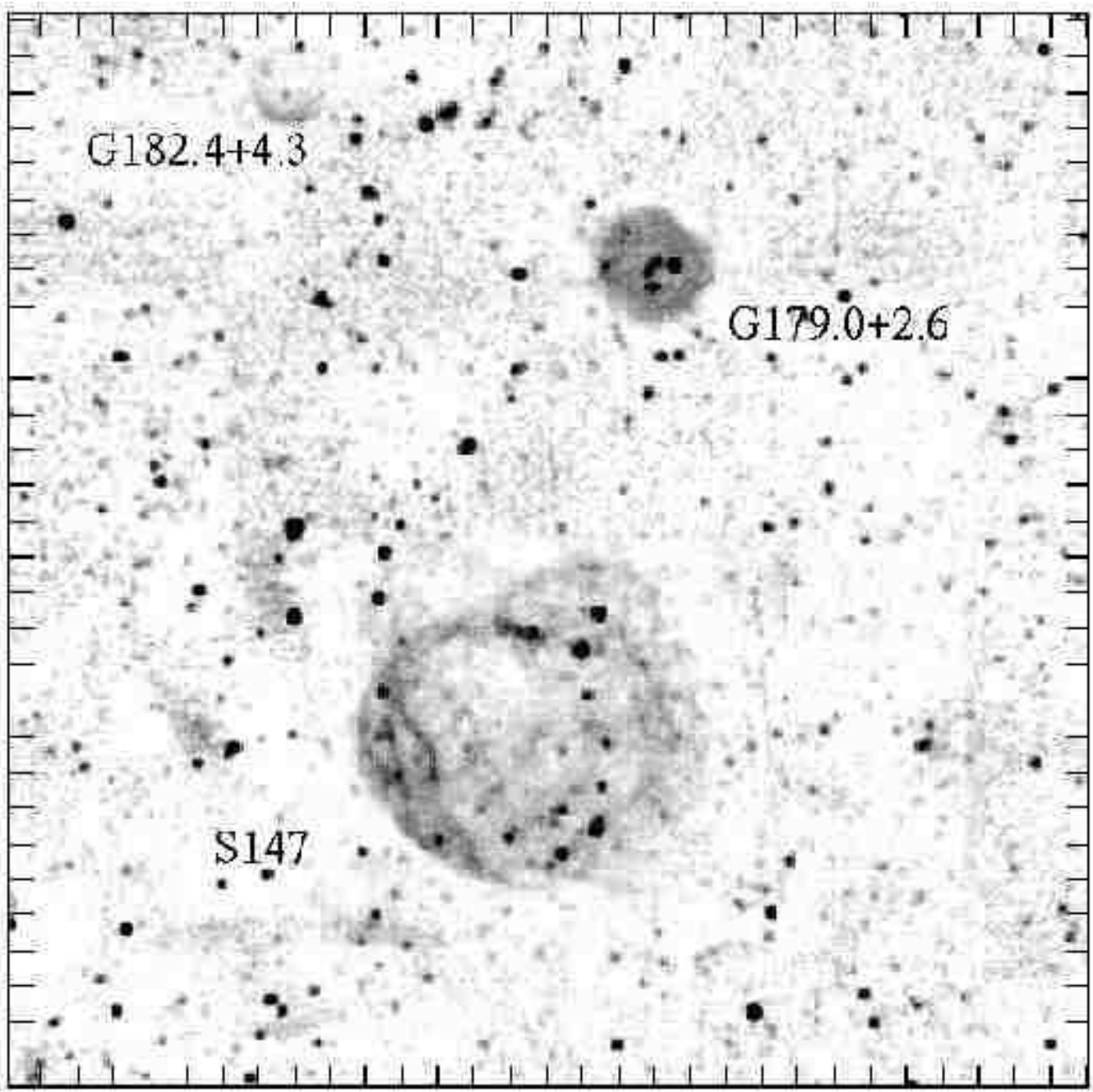}
\caption{Section towards the anticentre extracted from the Effelsberg
11~cm survey. The well studied SNR S147 and the two SNRs identified
from the 11~cm survey are indicated.
\label{image}}
\end{figure*}

\section{SNR catalogues, statistics and completeness}

There are two useful SNR compilations accessible through the web. One
is David Green's catalogue ({\tt
http://www.mrao.cam.ac.uk/surveys/snrs}),
where in its December 2001 version data and references for 231 SNRs
are compiled including information also from optical, infrared and
X-ray observations. Another useful database is that maintained by
Sergei Trushkin ({\tt http://cats.sao.com}) with radio flux densities,
spectra and references. A limited number of SNRs is seen at optical,
infrared and X-ray wavelength. Increased sensitivity resulted in more
detections. In particular in the X-ray range based on the ROSAT all-sky
survey a large number of SNR candidates have been identified
(Busser \cite{bus}). Thus it seems possible that the present fraction
of 30\% of SNRs showing up at X-rays will be increased, although the
low integration times in the ROSAT survey, its relatively soft energy
range and significant absorption for objects at larger distances in the
Galactic plane are limitations. Meanwhile for some cases subsequent
radio observations could prove the SN origin of ROSAT candidates (see
Schaudel et al., this volume).

The data in the catalogues are either from dedicated studies or from
radio continuum surveys. To what extent are the collected data complete?
From statistics about 40 SN should have occurred in the Galaxy within
the last two thousand years, although just eight are detected. As
expected the two strongest SNRs, which are also the strongest sources
in the sky, are Cas~A and Tau~A from A.D.1680 and A.D.1054,
respectively. It is unlikely that the present sensitivity of radio
surveys is insufficient to detect such sources throughout the Galaxy
when exploding in a typical interstellar environment. Certainly massive
stars are clustered and in case they explode within very dense molecular
clouds the lifetime of SNRs is reduced substantially. Compact
\ion{H}{ii} regions have a similar spectrum compared to plerions 
(as discussed in section 7), although the thermal emission from \ion{H}{ii} 
regions is unpolarized. 
However, confusion problems may be present.

For elder SNRs, which are expected to be intrinsically fainter than
young SNRs, the fraction of undetected SNRs may be similar or even
larger. From the 231 SNRs listed by Green just 9 SNRs are pure
plerions and 23 are classified as combined-type objects, e.g. a
shell-type SNR with a synchrotron nebula inside being powered by the
neutron star. These two groups make up about 15\% of all known SNRs,
while from SN events about 80\% are expected to leave a neutron star
and their remnants should be plerions or combined-type SNRs. Beside the
selection effects due to very dense material, which largely reduces
the SNR lifetime, a number of additional effects have to be considered,
which make the available SNR catalogues much likely rather incomplete.
The blast wave of SNRs expanding in a hot and thin interstellar medium
may reach quite a large size until enough material is swept up to
interact with. Faint objects with a diameter of several degrees are difficult
to identify in practice, even when offset from the Galactic plane. The
task is even more difficult when a regular shell is distorted by
fluctuations in the interstellar medium, which are quite likely to be
present. In the Galactic plane itself, with a high concentration of
sources and diffuse emission, confusion effects are severe and
sufficient high intensity and/or an undistorted morphology of the SNR
is required to be detected. High depolarization at lower
frequencies limits the search for polarized Galactic sources in addition.

\section{Identification of SNRs}

Radio continuum surveys are the main source for identifying new SNRs.
The Effelsberg surveys of the Galactic plane at 1.4~GHz and 2.7~GHz
wavelength were quite successful. References and the data are
accessible through the web: {\tt
http://www.mpifr-bonn.mpg.de/survey.html}. Figure~1 shows a section
from the 2.7~GHz survey towards the Galactic anti-center, where the
well-known SNR S147 is located, which shows a rather similar
morphology at radio and optical wavelength. Two fainter SNRs as
indicated in Fig.~1 could be identified (F\"urst \& Reich\
\cite{fue86a}, Kothes et al.\ \cite{koth98}). Also the more recent
Molonglo survey at 860~MHz covering the southern Galactic plane led to
a large number of new identifications (Whiteoak \& Green \cite{white}).
Of course, the newly detected objects are on average fainter than
previously known sources. Surveys at arcmin angular resolution are not
able to identify very compact, barely resolved sources and synthesis
telescope surveys are required like the VLA 1.4~GHz NVSS (Condon
et al.\ \cite{CBS89}) or the 408~MHz and 1.4~GHz CGPS survey carried out
at DRAO (Taylor et al.\ \cite{taylor}). Also the 327~MHz Westerbork
survey needs to be mentioned in that context (Taylor et al.\
\cite{tay96}). However, synthesis telescope surveys have the disadvantage
of being insensitive to large-scale structures and despite their low
confusion limit are often not able to measure extended objects reliably.

Morphology, e.g. shell-type sources, linear polarization and non-thermal
spectral indices are the classical tools to identify new SNRs. The
technique of comparing radio maps from the Effelsberg surveys with
those from the IRAS 60${\mu}$m or 100${\mu}$m surveys, which have
similar angular resolutions, was quite successful in identifying new
SNRs, since the ratio of infrared to radio flux for \ion{H}{ii} regions
is about an order of magnitude stronger than that for SNRs (F\"urst et
al.\ \cite{fue87}). Although significant variations in the infrared
to radio ratio of \ion{H}{ii} regions are evident, there is a clear
distinction compared to SNRs, which is in addition independent of
the type of SNR.

\section{Shell-type SNRs}

The vast majority of identified SNRs are of shell-type. This
morphology is expected when releasing a strong blast wave with an
initial expansion velocity of the order of 20000~km/sec. An estimated
energy of about $10^{51}$~ergs is released into the interstellar space.
In his classical paper Woltjer (\cite{wol72}) described the following
phases of the evolution of a shell-type SNR with an initial energy
$\rm E_{o} = 10^{51}$~ergs into a homogeneous interstellar medium with
density $\rm n_{o}\, [cm^{-3}]$.

Free expansion: the blast wave expands with its initial velocity and
the radius of the SNR increases linear with time. This phase ends until
about 1 to 10 M$_{\odot}$ of material are swept up from the ambient
medium in the SNR shell.

Adiabatic expansion or Sedov phase: the energy of the SNR is
conserved. The radius $\rm R_{s}\, [pc]$ increases as $\rm R_{s} =
14~(E_{o}/n_{o})^{1/5}~t^{2/5}$, with t in units of $10^{4}$ years.
The expansion velocity of the SNR decreases accordingly. The density in
the shell is $\rm n_{s} = 4~n_{o}$ and its thickness is about 10\% of
$\rm R_{s}$. This phase of SNR evolution ends when about half of its
initial energy is released. According to Cox (\cite{cox}) this phase
ends when the age t of the SNR  is $\rm t_{rad} =
4.3~E_{o}^{4/17}~n_{o}^{-9/17}$, its radius is $\rm R_{rad} =
23.1~E_{o}^{5/17}~n_{o}^{-7/17}$ and its expansion velocity $\rm
v_{rad}~[km\, s^{-1}] = 210~E_{o}^{1/17}~n_{o}^{2/17}$.

Finally the SNR enters the radiative phase until its initial energy
$\rm E_{o}$ is released to the ISM. In this phase radius and expansion
velocity are $\rm R_{s} = R_{rad}~(8 t/5 t_{rad} - 3/5)^{1/4}$ and
$\rm v_{s} = v_{rad}~(R_{rad}/R_{s})^{3}$. The SNR merges with the
interstellar medium at $\rm t_{c} = 1.23~t_{rad}$ and $\rm R_{c} =
1.05~R_{rad}$.

Although this evolution scenario describes the phases of SNR evolution
in principle correctly as proven by observations, it is certainly
a simplification. The assumption of a constant density of the ISM
during all phases of SNR evolution is not realistic. The effect of the
stellar wind from the massive progenitor star modifies the ISM and the
young SNR evolves in a quite thin medium and later will encounter the
dense material previously swept up by the stellar wind (McKee et al.\
\cite{mckee}). Inhomogeneities of the ISM may distort a regular SNR
shell, gradients in density may cause an elliptical shape and low
density bubbles may lead to ``break-out'' phenomena. Also the structure
and regularity of the ambient interstellar magnetic field shapes SNRs
(van der Laan\ \cite{laan}) and barrel-shaped structures might result.
These in turn indicate the direction and homogeneity of the ambient
magnetic field.

\subsection{Synchrotron emission}

SNRs are strong emitter of synchrotron emission and their spectra can
well be described by a power law $\rm S \sim \nu^{\alpha}$, with S the
observed flux density and $\nu$ the observing frequency. The SNR
blast wave sweeps up and compresses the ambient magnetic field $\rm B_{0}$ that an
enhanced volume emissivity $\rm (B/B{0})^{1-2{\alpha}}$ results. In this case
the spectrum of a SNR reflects the cosmic ray electron spectrum, which
steepens towards higher energies, and the spectral index ${\alpha}$ varies accordingly. The
strong magnetic field shifts spectral variations towards higher
frequencies. In addition the strong shock at the SNR blast wave also
induces turbulences, which are able to accelerate particles up to very
high energies. Known processes are Fermi~I and Fermi~II
acceleration.

The majority of SNRs are in the adiabatic phase and have spectra
around $\alpha =-0.5$. This is expected for a strong shock
and a compression factor of 4. Their magnetic field is tangential.
Young (historical) SNRs show steeper spectra ($\alpha =-0.6$ to $-$0.8)
on average. Their magnetic field is radial.

\subsection{Very young SNRs}

\begin{figure}
\includegraphics[width=8.8cm,angle=180,clip]{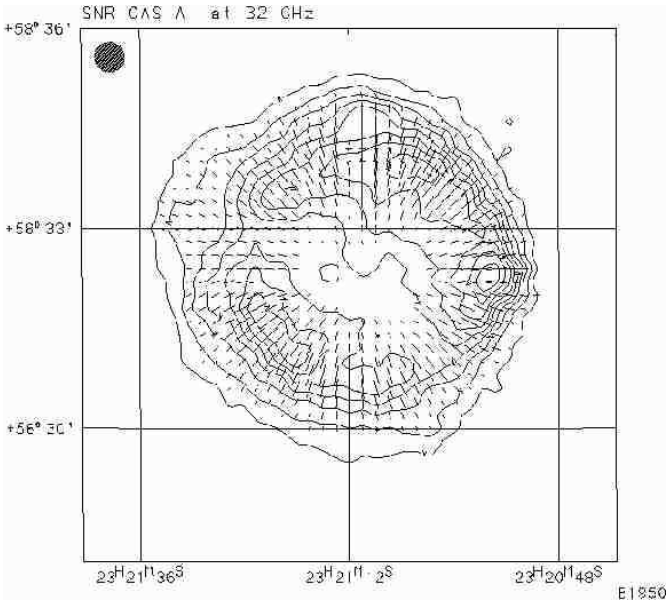}
\caption{Cas~A observed with the Effelsberg 100-m telescope with
$26\arcsec$ angular resolution. Polarization bars in B-field direction
are superimposed.
\label{image}}
\end{figure}

Sensitivities in the mJy range are required to observe SN in nearby
galaxies and a number of SN events have been measured in recent years.
A quite well-studied example is SN1993J (a IIb type) in M81. The first
radio detection was about 4 days after the explosion, when about
0.8~mJy were seen at 22~GHz. A steady rise of the flux density to
about 1.7~mJy about 50 days later was noted. Afterwards the flux
density slowly declined. At lower frequencies radio emission was seen
delayed as is expected for an optically thick source. The optical
thickness decreases with expansion. P\'erez-Torres et al. (\cite{per})
fitted the observed radio light curves for a number of frequencies and
in fact found evidence for synchrotron self-absorption, however,
external free-free absorption was required in addition.

Today's high sensitivity VLBI technique allows to follow also the
spatial evolution of young SNRs. Marcaide et al. (\cite{mar})
analyzed a sequence of VLBI images for the first 42 months of SN1993J
and found an increase of its radius with time as $\rm R \sim
t^{0.89\pm0.03}$, which is close to -- although definitely not
identical with -- a linear expansion of the blast wave. Although
the images show some intensity variations in the shell Marcaide et al.
(\cite{mar}) suggest some residual effect in the image restoration, but
take its almost circular shape as evidence for a symmetric explosion.

\subsection{Historical SNRs}

\begin{figure}
\includegraphics[width=9cm,angle=180,clip]{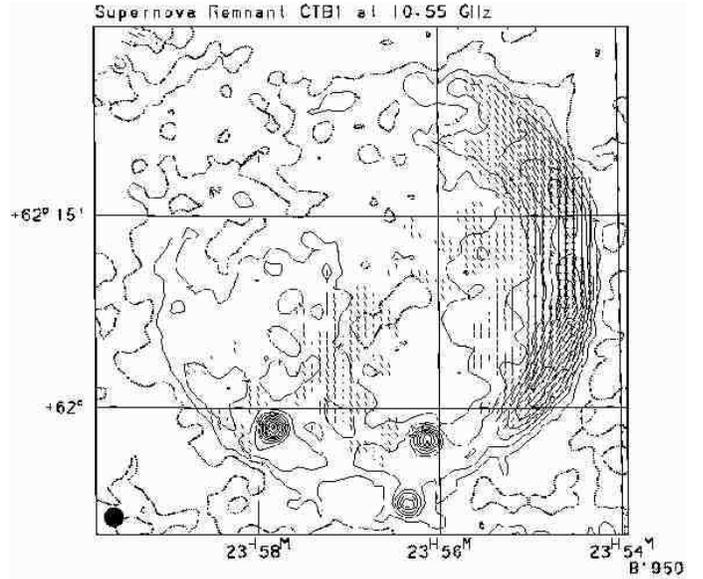}
\caption{SNR CTB~1 observed at 10.55~GHz (HPBW~=~69\arcsec ) with the
Effelsberg 100-m telescope. Polarization bars are in B-field
direction assuming negligible Faraday rotation.
\label{image}}
\end{figure}

\begin{figure}
\includegraphics[width=8.8cm,angle=180,clip]{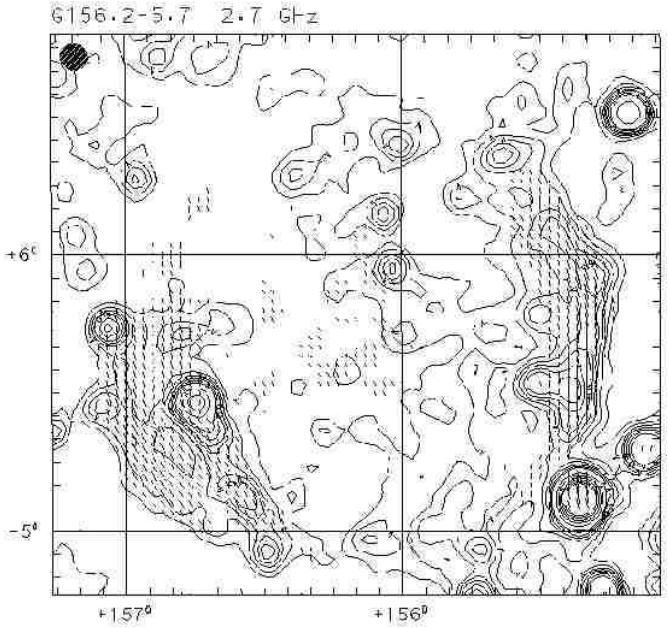}
\caption{SNR G156.2+5.7 in the Galactic coordinate system. The
Effelsberg 2.7~GHz image is shown convolved to $6\arcmin$ (HPBW)
with polarization bars in B-field direction for the (very likely) case
of low Faraday rotation (Reich et al.\ \cite{rei92}).
\label{image}}
\end{figure}

Until recently historical SNRs resulting from SN events within the
last 2000 years were mostly identified via optical flaring
events noted in ancient records. There are the shell-type sources
SN~1006, Tycho and Kepler. The plerions Tau~A and 3C58 and the
combined-type SNR G11.2$-$0.3. In addition the shell-type SNRs Cas~A and
RXJ0852.0$-$4622 belong to this group, where the detection of
the short living $\rm Ti^{44}$ with a mean lifetime of 90.4~years
proves their young age. Cas~A was previously believed to originate
from a SN event in 1680, but no clear optical record exists.
RXJ0852.0$-$4622 is not an outstanding radio source, but has been
clearly detected in a follow-up search in a highly confused region of
the sky (Duncan et al.\ \cite{dun}).

In Figure~2 we show a 32~GHz image of Cas~A obtained with the Effelsberg
100-m telescope. Although its spectrum is quite steep with a spectral
index of $\alpha = -0.77$, it is nevertheless one of the strongest
sources at mm-wavelength. As expected its magnetic field is almost
perfectly radial. Along its outer northern rim Gotthelf et al.
(\cite{gott}) recently found evidence for a weak tangential component
at its outer shock, indicating that Cas~A is leaving free expansion
and begins to enter the Sedov phase. The considerable total intensity
variations along the shell indicate some influence of an inhomogeneous
interacting medium.

\begin{figure*}
 \centering
 \begin{minipage}[c]{14.5cm}
\includegraphics[width=13.5cm,angle=90,clip]{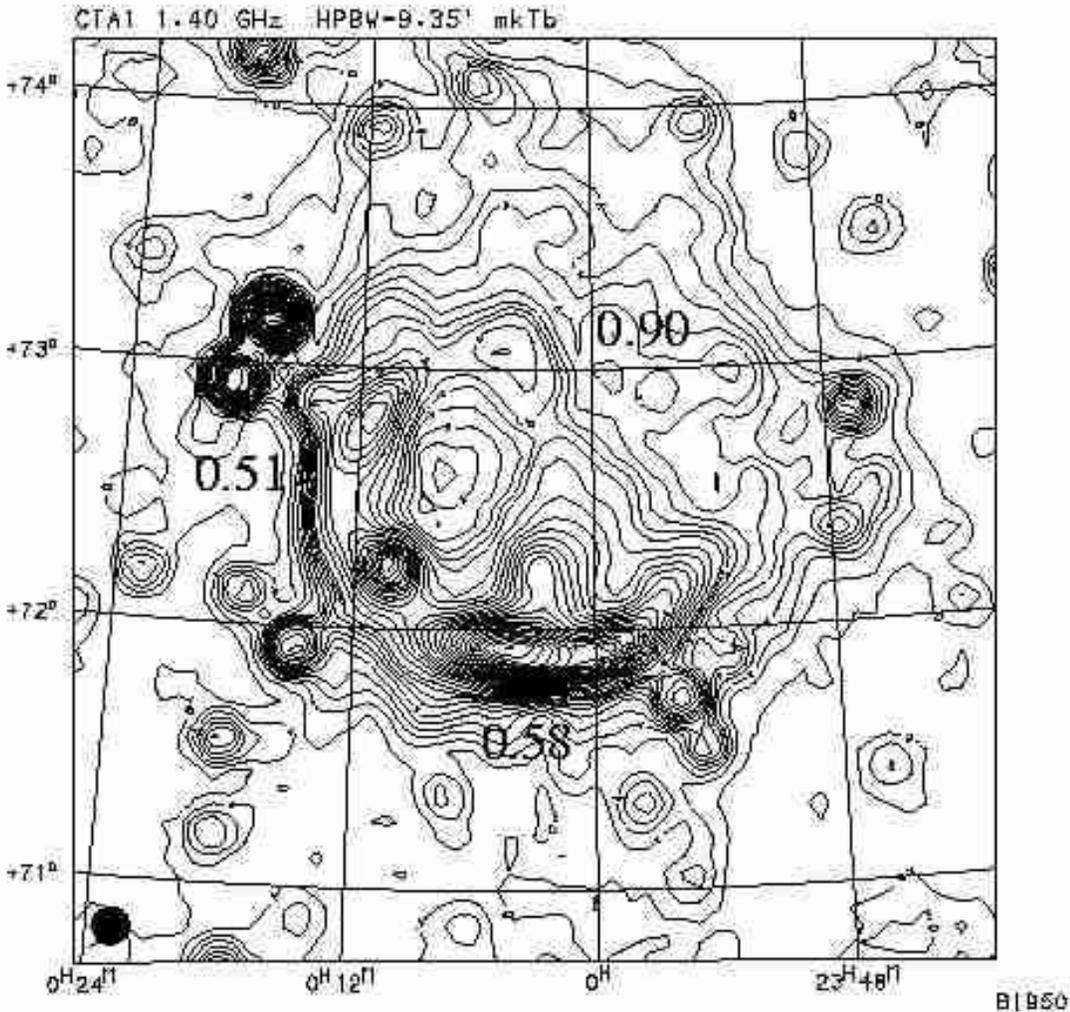}
\end{minipage}%
\hfill
 \begin{minipage}[c]{3.5cm}
 \centering
\caption{Effelsberg map of CTA~1 at 1.4~GHz (Pinault et al.
(\cite{pin97}). The map is at 9\farcm 35 angular resolution
(HPBW). The (absolute) spec\-tral indices for the shell and the
`break-out' region are indicated.} \label{image}
\end{minipage}
\end{figure*}

\subsection{Adiabatic phase}

As mentioned earlier most of the detected SNRs are in the adiabatic
phase with a spectral index close to $\alpha =-0.5$. Figure~3
shows an image of CTB~1 at 10.55~GHz (Schmidt et al.\ \cite{Schmidt})
with a spectral index of $\alpha =-0.58$. This SNR is located at a
distance of about 2~kpc as found from associated neutral hydrogen
(Landecker et al.\ \cite{tl82}) and has a diameter of about 21~pc. A
well-defined shell with a tangential magnetic field is seen along its
western periphery. In the south and south-east the shell is still visible,
but less pronounced and is almost missing in the north-eastern
quadrant. Likely the ISM density is lower in this region, which
is supported by X-ray emission extending beyond the circular shell
periphery and indicating some kind of `break-out', where the
blast wave expands faster and less material is swept up.

Another example of the `break-out' phenomenon is indicated for CTA~1,
a high latitude SNR at an estimated distance of about 1.4~kpc.
CTA~1 has a flat-spectrum non-thermal X-ray core, in addition thermal
X-ray emission related to the blast wave of the SNR and seems to be
associated to the $\gamma$-ray source J000702+7302.9 (Slane et al.\
\cite{sla97}). Thus CTA~1 has features not in common to the majority of
adiabatic SNRs and may be classified as composite SNR from its X-ray
morphology. However, details of its radio emission can well be studied
due to its location out of the plane. A 1.4~GHz map (Pineault et al.\
\cite{pin97}) is displayed in Fig.~5, where also the spectral indices
in the shell and in the `break-out' region are indicated. There are
significant spectral differences of the order of $\alpha =0.25$
between the intense shells and the diffuse emission extending towards
the north-west. Pineault et al. (\cite{pin97} discuss this effect in
terms of Fermi acceleration in shocks and found it possible to account
the steepening of the spectrum by expansion of the SNR into a low
density region with small compression ratio.

SNRs in the adiabatic phase show a number of effects from their
interaction with the ISM, thus reflecting the properties of their
environment, which might be additionally largely shaped by the effect
of the stellar wind from the progenitor star. Some of these effects are
briefly discussed in turn:

The bi-symmetric morphology and the tangential magnetic field in the
shell has been used to derive the orientation of the interstellar
magnetic field. Objects with the most clear morphology have the most
uniform environment. The inclination of the magnetic field to the
Galactic plane is small in most cases as expected for a magnetic field
orientation along the spiral arms. However, some SNRs show larger
deviations. The highly polarized low surface-brightness SNR G156.2+5.7
(Fig.~4), for instance, is inclined by about $70\degr$. Although
its distance is not well constrained (1~kpc to 3~kpc), its size ranges
between 33~pc and 95~pc and therefore the inferred regular magnetic
field orientation holds for a region similar or larger in size.

Rotation measure (RM) observations reveal the direction of the
magnetic field $B_{||}$ in the SNR shell of thickness l along the line
of sight. RM is calculated: $\rm RM\, [rad/m^{2}] = 0.81~n_{e}\,
[cm^{-3}]~B_{||}\,[{\mu}G]~l\,[pc]$, where $\rm n_{e}$ is the thermal
electron density. Polarization angles $\phi$ at three frequencies are
required to get an RM without ambiguity and with the correct sign,
where $\rm \phi\,[rad] = RM\, [rad/m^{2}]~\lambda\,[m]^{2}$. Again,
in most cases the magnetic field direction agrees with the global
field as inferred from averaging RMs of extragalactic sources and
pulsars. But there are exceptions as already noted by F\"urst \& Reich
(\cite{fue90}). Recently, Uyan{\i}ker et al. (\cite{bu01}) made a
detailed analysis of the polarization data of G93.7$-$0.3 (CTB104A),
where a RM gradient across the source was measured. This does not
agree with the direction of the global magnetic field at the distance
of the SNR of about 1.5~kpc, but agrees with a similar anomaly in the
RM data of pulsars. The size of the SNR of about 35~pc is quite likely
a rather conservative lower limit of this RM anomaly.

SNRs are supposed to trigger star formation, but just few observations
have been made to settle that topic. Molecular clouds interacting
with SNRs are the most promising cases where the blast wave may lead
to large density enhancements. In a study of G54.4$-$0.3 Junkes et al.
(\cite{jun92}) found evidence for a massive molecular shell
surrounding the SNR. While the average density is about $\rm
30\,[cm^{-3}]$, a number of higher density clumps (density $\rm
100-300\,[cm^{-3}]$ are embedded. A detailed analysis showed that
the SNR blast wave is likely not responsible, but the stellar wind of
the progenitor star has formed the molecular shell.

Shocked molecular gas in interacting clouds is direct evidence for the
interaction of the blast wave. The prototype SNR for associated shocked
molecular gas is IC443, but also other SNRs show shocked gas. The
association of $\gamma$-ray sources with interacting SNRs is clearly
established and reflects the absorption of high energy cosmic rays just
accelerated in the SNR shock-front with dense gas.

For a larger number of SNRs OH maser emission was clearly detected.
These observations indicate small condensations within the SNR shell,
where the magnetic field strength ranges up to about 0.5~mGauss
(Claussen et al.\ \cite{claus}).

\subsection{Evolved SNRs}

When the SNR enters the radiative phase its shell is highly
compressed, although its interior remains hot as its cooling
time is longer and therefore is still detectable by X-ray
observations. S147 (see Fig.~1) is an example of this class
of SNRs. Its filamentary radio morphology is also seen optically with
a perfect correspondence for many substructures (F\"urst \& Reich\
\cite{fue86b}). Evolved SNRs become faint and their surface brightness
$\rm \Sigma~[W\, m^{-2}\, Hz^{-1}\, sr^{-1}]$, which is calculated from
$\rm \Sigma = 1.5~10^{-19}\, S\,[Jy]/\Theta\, [']^{2}$, becomes rather
low.  Here S is the observed flux density and $\Theta$ is the measured
diameter of the SNR. For convenience $\Sigma$ is at most calculated for
1~GHz flux densities. $\Sigma$ is distance-independent and the
$\Sigma$--diameter relation was used for distance estimates, assuming
the density of the ISM to be fairly uniform. This turns out to be not
relevant, and in consequence large density differences reflect in
$\Sigma$ at a certain diameter. However, Berkhuijsen (\cite{berk})
combined Galactic SNR data with well-defined distances and
surface-brightness data from SNRs in nearby galaxies. She concluded
that for a given $\Sigma$ a maximum diameter can be inferred.
Accordingly S147 has a rather low surface brightness of $\rm
\Sigma_{1GHz} = 4~10^{-22}\, [W m^{-2}\, Hz^{-1}\, sr^{-1}]$.
However, the surface brightness of G179.3+2.6 ($\rm \Sigma_{1GHz} =
2~10^{-22}\, [W\, m^{-2}\, Hz^{-1}\, sr^{-1}]$) and G182.4+4.3 ($\rm
\Sigma_{1GHz} = 7.5~10^{-23}\, [W\, m^{-2}\, Hz^{-1}\, sr^{-1}]$) (see
Fig.~1) is even lower, although they do not show the properties of
evolved SNRs like S147 (F\"urst \& Reich\ \cite{fue86a}, Kothes et al.\
\cite{koth98}). It is most likely that G182.4+4.3 is in the adiabatic
phase of expansion. Figure~6 displays a 4.85~GHz Effelsberg image
showing a tangential magnetic field and also its spectral index of
$\alpha = -0.42\pm0.10$ is in agreement with SNRs in the adiabatic
phase. It is remarkable that its size of about 45~pc is not much
smaller compared to that of S147 (42~pc to 82~pc), despite the large
systematic uncertainties for S147.

Evolved SNRs with a compressed magnetic field where no particle
acceleration takes place any longer reflect the Galactic electron
spectrum with its known bends in their spectra, although the
compressed magnetic field shifts the observing frequencies.
S147 shows all these characteristics (F\"urst \& Reich\ \cite{fue86b}).
Its integrated flux-density spectrum bends at about 1~GHz. However, the
spectrum of its filaments is flatter than that of the diffuse
emission. Due to the more compressed magnetic field in the filaments
radiation from low-energy electrons with a flatter energy spectrum are
seen. In the diffuse part of the source the lower magnetic field
strength traces emission from electrons with higher energies, which
have a steeper energy spectrum.

Despite the remaining uncertainty in classifying a SNR from its low
surface brightness, more identifications of these weak sources
are of interest to understand SNR evolution and their impact on the
ISM. Faint large-diameter objects are filtered out by synthesis
telescope observations, while the sensitivity of single-dish
observations is limited by confusion of faint unresolved
extragalactic sources in their larger beams. An improvement seems
possible, when deep-source surveys and single-dish surveys exist at
the same frequency and the source contribution can be well subtracted
from the single-dish map. An example of such an attempt is shown in
Fig.~7, where the noise in the original 1.4~GHz Effelsberg map
(Uyan{\i}ker et al.\ \cite{bu99}) was significantly lowered
by subtracting all compact sources visible in the NVSS (Condon et al.\
\cite{CBS89}). Two shells or partial shells are visible in Fig.~7 with
diameter of about $3\degr$ or $1\fdg 6$. The surface brightness is
estimated to 
about $\rm \Sigma_{1GHz} = 2.5~10^{-23}\, [W\, m^{-2}\, Hz^{-1}\,
sr^{-1}]$ and $\rm \Sigma_{1GHz} = 3.8~10^{-23}\, [W\, m^{-2}\, Hz^{-1}\,
sr^{-1}]$.

\begin{figure}
\includegraphics[width=8cm,angle=-90,clip]{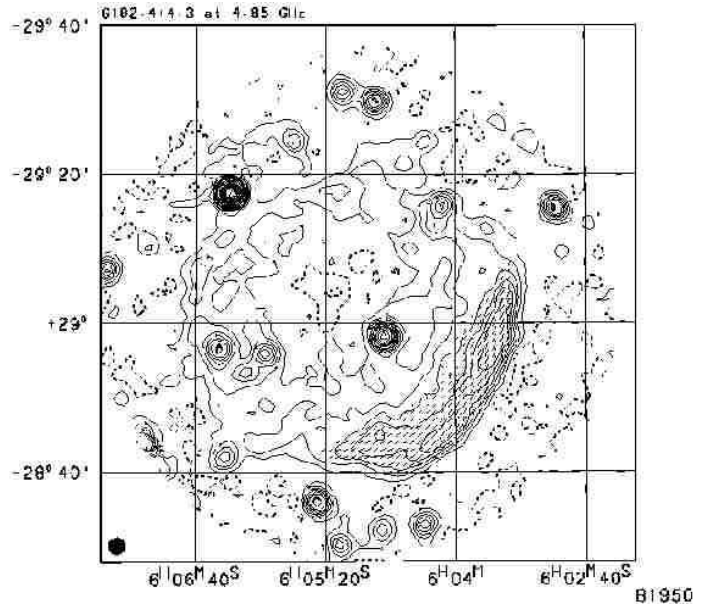}
\caption{G182.4+4.3 at 4.85~GHz convolved to a HPBW of $3\arcmin$
(Kothes et al.\ \cite{koth98}). Polarization bars show the
B-field direction in case of low Faraday rotation.
\label{image}}
\end{figure}

\begin{center}
\begin{figure*}
\includegraphics[width=17.5cm,clip]{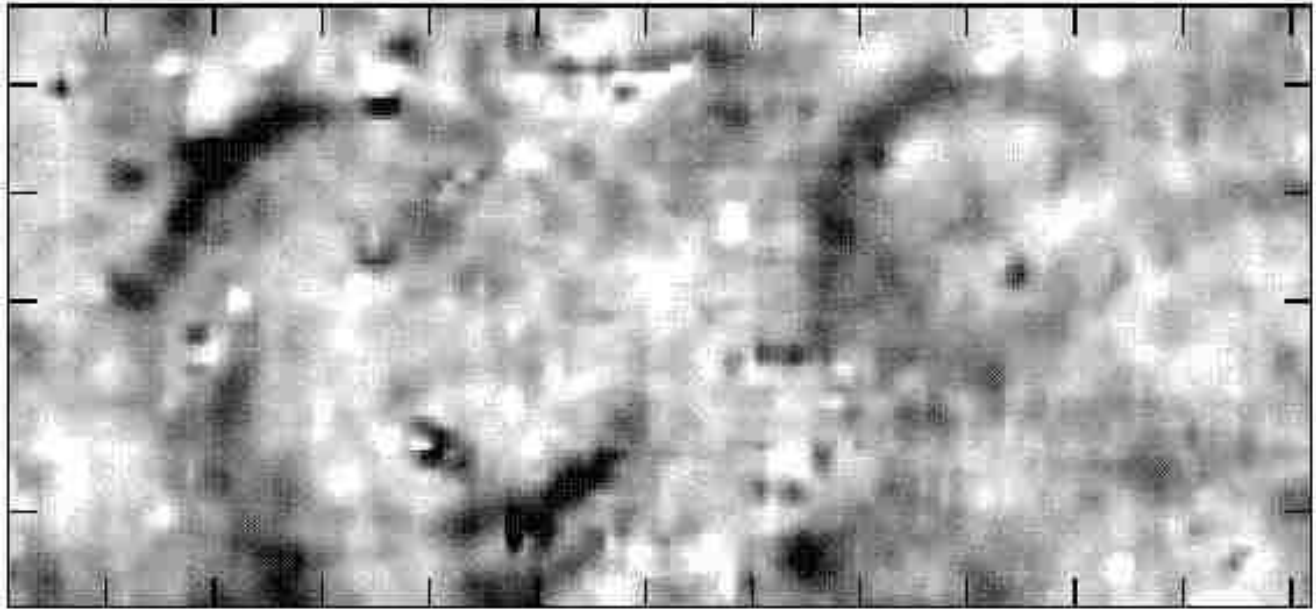}
\caption{Two low surface-brightness SNR candidates at 1.4~GHz towards
the Galactic anticentre direction (Soberski \& Reich, in prep.). The
map is a section from the Effelsberg 1.4~GHz Galactic Medium Latitude
Survey (extracted from Fig.~12 of Uyan{\i}ker et al.\ \cite{bu99})
with NVSS sources (Condon et al.\ \cite{CBS89}) subtracted. The map
size is about $8\degr \times 4\fdg 7$.
\label{image6}}
\end{figure*}
\end{center}

\section{Pulsar--SNR associations}

Although Pulsar~(PSR)--SNR associations are expected to be quite
common, as they result from a core collapse event, the number of
identifications is still small. Before 1980 just Tau~A and the Vela SNR
were known to have an associated PSR. Since then a number of new detections
were made. Until now about 20 associations were identified based on
sensitive X-ray as well as on radio observations. Proven associations
show young neutron stars, allowing to determine their properties
including beaming and birth rate. Better age and distance estimates
become possible. PSR distances are inferred from \ion{H}{i} absorption,
their RM and dispersion measure (DM), while SNR distances are estimated
from \ion{H}{i} absorption, associated \ion{H}{i} clouds or molecular
emission structures.

\begin{figure}
\includegraphics[width=8.8cm,clip]{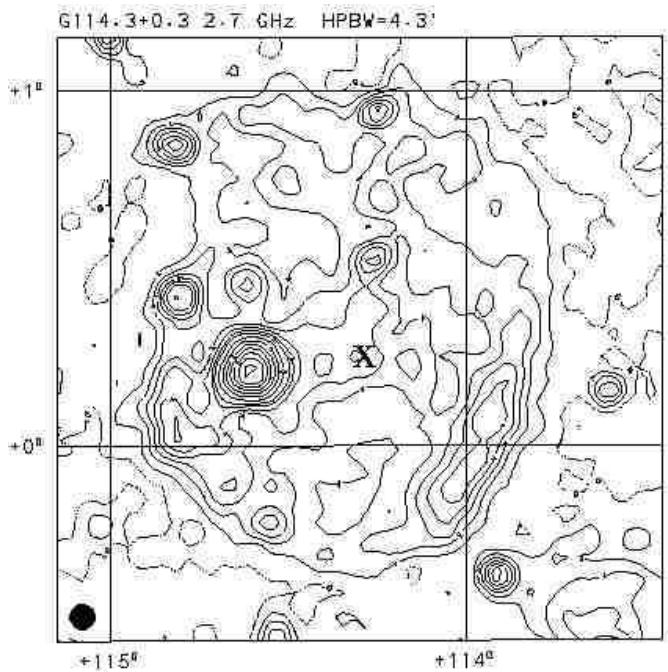}
\caption{Image of SNR G114.3+0.3 extracted from the Effelsberg 2.7~GHz
Galactic plane survey with indicated position of PSR2334+61 close to
its center. The compact sources in the field are extragalactic
except the bright source east of the SNR center, which is an unrelated
\ion{H}{ii} region.
\label{image}}
\end{figure}

\begin{figure}
\includegraphics[width=8cm,angle=-90,clip]{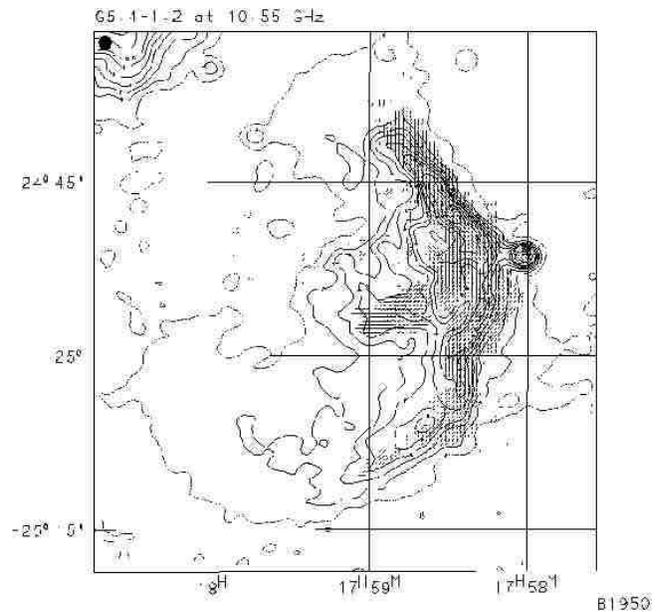}
\caption{Effelsberg image of G5.4$-$1.2 with polarization bars along the
B-field direction. The PSR B1757$-$24 is located just west of its wind
nebula, which is located just outside the SNR shell. Its passage
causes a clearly visible shell distortion.
\label{image}}
\end{figure}

S147, as discussed in the previous section as an example of an evolved
SNR, has a 143~msec PSR detected in 1996 at about $40\arcmin$ from the
center of the SNR (Anderson et al.\ \cite{and}). Another example is
G114.3+0.3/PSR2334+61 as shown in Fig.~8. This low surface brightness
object has a 0.5~sec PSR close to its center. From combining all
available information F\"urst et al. (\cite{fue93}) concluded that a
distance of 2.2~kpc and an age of $2~10^{4}$~years is most likely for
this association. The inferred transverse velocity of the PSR from the
geometric center of the SNR is about 125~km/s, well within the range of PSR
velocities.

A rather interesting case is the association SNR G5.4$-$1.2/PSR
B1757$-$24. In Figure~9 an image at 10.55~GHz is displayed, where the
PSR has overtaken the SNR shell with high velocity. Its cometary tail
powered by the pulsar wind is directed towards the SNR center. Gaensler
\& Frail (\cite{gaen00}) were able to limit the proper motion of the
PSR from multi-epoch VLA measurements. This results in an age estimate
3 to 10 times larger than its `characteristic age', which relates the
PSR age to the measured derivative of the rotational frequency.

Recently Kothes et al. (\cite{ko00}) observed the association
G106.3+2.7/PSR J2229+6114 and found a highly polarized flat-spectrum
structure, the `boomerang' ahead of the PSR. This bow-shock structure
results from an interaction of the PSR wind with surrounding cold and
dense material, likely previously swept up from the stellar wind of
the progenitor star of G106.3+2.7.

Kothes (\cite{ko98}) found a very close correlation between $\rm
\Sigma_{1GHz}$ of PSR nebula, their diameter and the PSR energy loss
rate $\rm \dot{E}$: $\rm \Sigma_{1GHz} \sim \dot{E}/D$. This correlation
holds for about four orders of magnitude of $\rm \Sigma_{1GHz}$ and
must be accounted for by evolutionary models. The correlation seems
also useful in predicting PSR parameters from the observed
properties of its synchrotron nebula.

\section{Plerions and combined-type SNRs}

Even in cases where the PSR cannot be detected, its wind of
relativistic particles may become pressure-confined and subsequently
form isotropically radiating nebula. This classifies the SNR as
resulting from a core collapse event. The identification of these
flat-spectrum cores within steep non-thermal emission from the SNR
shell requires low and high frequency observations to separate
both components. An example of this class of objects is G11.2$-$0.3,
believed to be the SNR from the A.D.~386 supernova. An image is
displayed in Fig.~10 (Kothes \& Reich\ \cite{ko2000}). Recently a
65~msec X-ray pulsar was detected by Torii et al. (\cite{tor}),
which was later proved by sensitive CHANDRA observations to be
precisely in the center of G11.2$-$0.3 (Kaspi et al.\ \cite{kas}).
Thus the supplying source of relativistic particles is identified.
Also in this case the characteristic age of about 24000~years
largely disagrees with its true age.

\begin{figure}
\includegraphics[width=9cm,angle=90,clip]{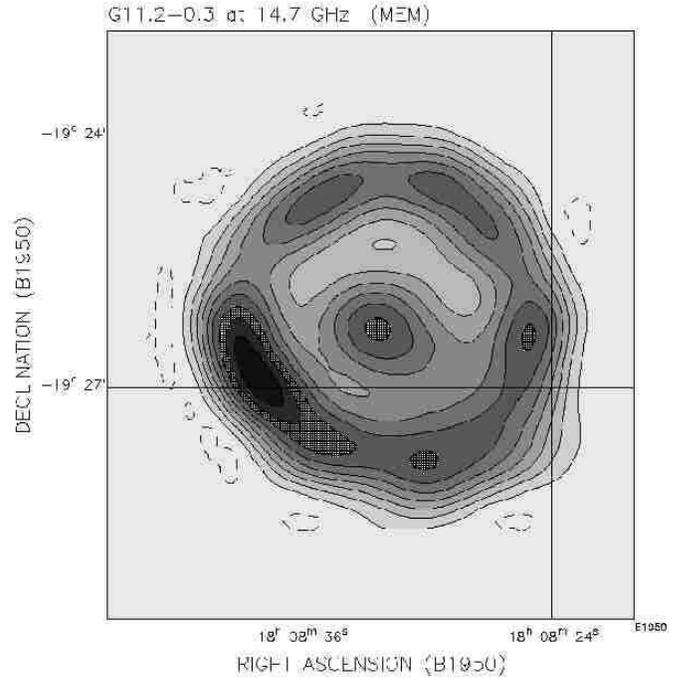}
\caption{G11.2$-$0.3 MEM image from Effelsberg observations (Kothes \&
Reich\ \cite{ko2000}) showing the flat-spectrum core inside the steeper
spectrum SNR shell.
\label{image}}
\end{figure}

Another somewhat different example of a combined-type SNR, where --
despite of dedicated searches -- so far no PSR has been identified, is
G18.95$-$1.1 as shown in Fig.~11. This object exhibits a number of
differences compared to G11.2$-$0.3 or other combined-type SNR, as it is
dominated by flat-spectrum centrally-peaked diffuse emission with
filaments running outwards from a central bar-like feature. Its shell
is only partial formed. F\"urst et al. (\cite{fue97}) compared the
Effelsberg 10.55~GHz radio image and the X-ray image obtained with
the PSPC of ROSAT. A remarkable similarity of the diffuse emission in
the radio and X-ray band was noted, suggesting that the synchrotron
emissivity and the thermal emissivity are closely related. In case of
pressure equilibrium it could be shown that the tangential magnetic
field component decreases radially much faster than the total magnetic
field strength. A wound-up magnetic field is suggested. For G18.95$-$1.1
and also other sources, a simple model of an isotropically expanding
PSR-driven synchrotron bubble seems insufficient to account for the
observations.

\begin{figure}
\includegraphics[width=8cm,angle=-90,clip]{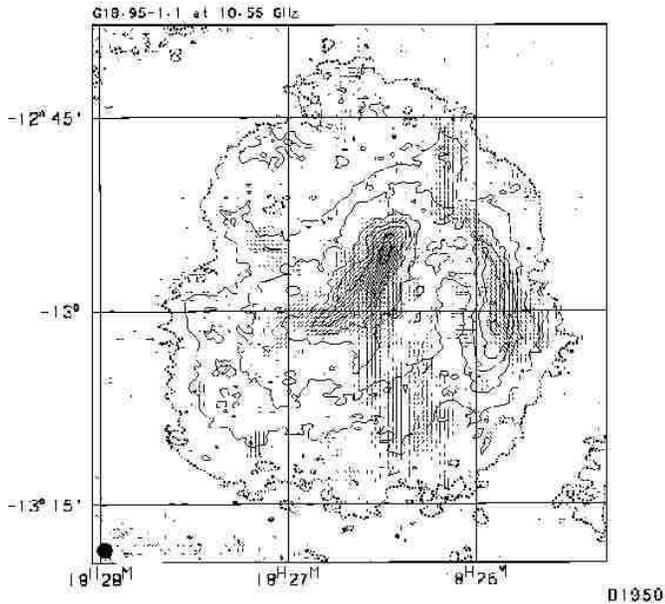}
\caption{The combined-type SNR G18.95$-$1.1 as observed with the
Effelsberg 100-m telescope. Polarization percentages up to 30\%
are seen. Polarization bars are along the B-field.
\label{image}}
\end{figure}

\begin{figure}
\includegraphics[width=7.5cm,angle=-90,clip]{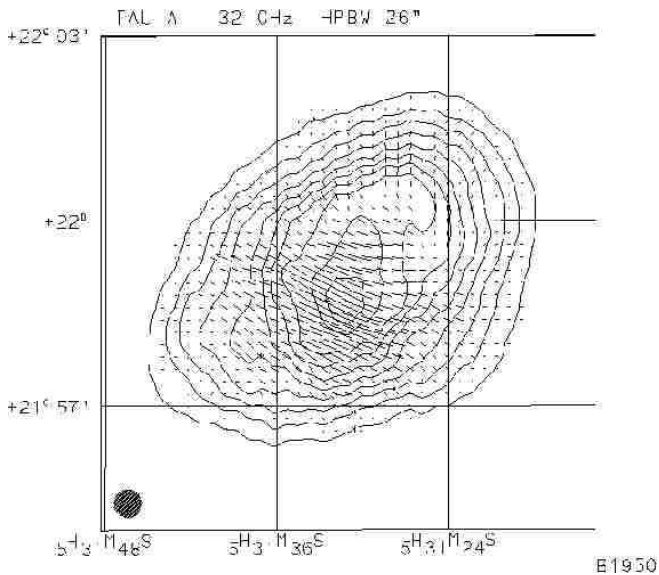}
\caption{Tau~A as observed with the Effelsberg 100-m telescope.
Polarization bars in B-field direction are superimposed.
\label{image}}
\end{figure}

\begin{figure}
\includegraphics[width=9.2cm,angle=90,clip]{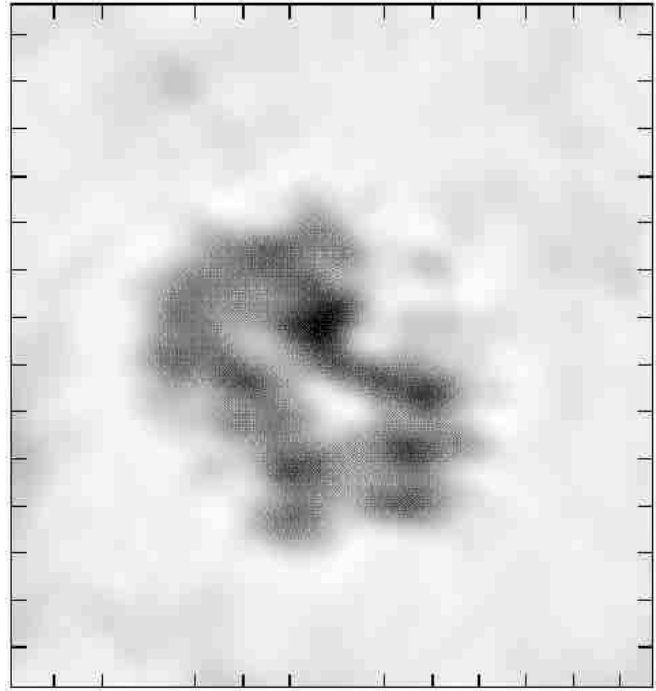}
\caption{Small-scale structure of G21.5$-$0.9 at 22.3~GHz as observed
with the Nobeyama Array. The map is convolved to $8\arcsec$ angular
resolution (HPBW) and shows an area of $2\farcm 3\times 2\farcm 5$. In
the center of the symmetrical double-lobe structure a compact X-ray
source is located.
\label{image}}
\end{figure}

\begin{figure}
\includegraphics[width=8.8cm,angle=180,clip]{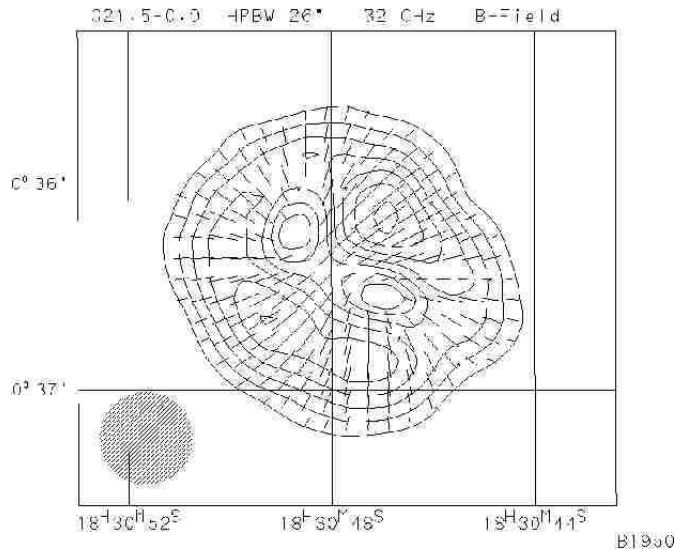}
\caption{Contours show the polarized 32~GHz emission of G21.5$-$0.9 from
Effelsberg observations. The radial B-field is shown by
the superimposed polarization bars.
\label{image}}
\end{figure}

Plerions or Crab-like SNRs, which show up as pure filled-center
objects, have Tau~A -- the Crab nebula -- as the prototype source. An
image of Tau~A is displayed in Fig.~12. Plerions emit flat-spectrum
synchrotron emission and are significantly polarized. Despite very deep
searches for some objects no radio shell could be identified
reflecting the impact of the blast wave from the SN explosion with the
ISM. Nevertheless plerions might be considered as a subgroup of
combined-type SNRs with quite bright flat-spectrum emission and
therefore their low-frequency spectra are better to determine compared
to the confused cores of combined-type SNRs. Woltjer et al.
(\cite{wol97}) discussed the properties of plerions in comparison to
Tau~A, which has a radio and X-ray luminosity of about two orders of
magnitude larger than all other members of the group. While Tau~A can
well be described by the PSR-powered expanding sphere model of Pacini
\& Salvati (\cite{pac73}), which correctly predicts a spectral
steepening of $\Delta\alpha = -0.5$ in the infrared due to synchrotron
aging, most of the other plerions have spectral breaks in the range of
20~GHz to 50~GHz. This is incompatible with synchrotron aging, and
Woltjer et al. (\cite{wol97}) considered cases of a non-standard
evolution of the PSR injection.

\begin{figure*}
\includegraphics[width=17.6cm,angle=180,clip]{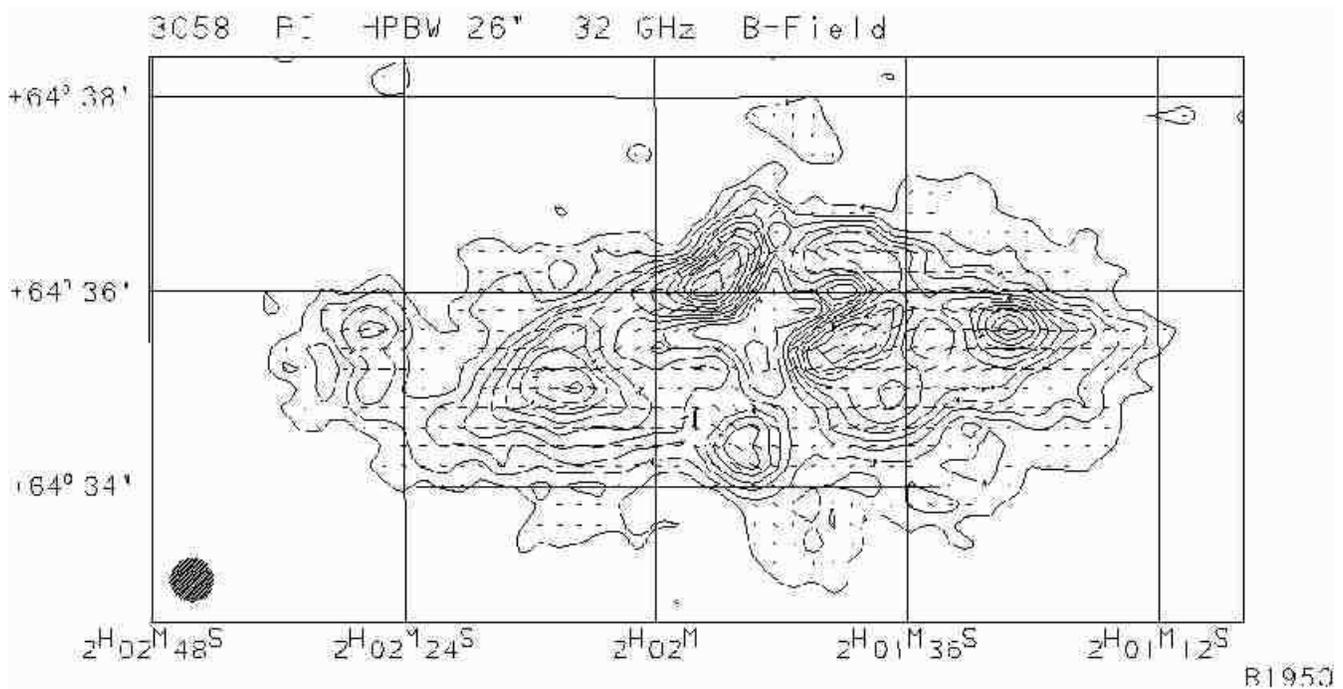}
\caption{Effelsberg image of the polarized intensity of 3C58 (Reich et
al.\ \cite{rei98}). Bars are shown in B-field direction}
\label{image}
\end{figure*}

A well-studied case of a plerion is G21.5$-$0.9, which was observed with
high angular resolution by F\"urst et al. (\cite{fue88}) using
the Nobeyama Array at 22.3~GHz. 80\% of its emission originates from a
diffuse slightly elliptical centrally peaked component and 20\% comes
from a double-cone structure as shown in Fig.~13. A PSR is not visible
in the radio, but a non-pulsating X-ray emitting compact source is
observed (Slane et al.\ \cite{sla00}, Warwick et al.\ \cite{war01}).
This source is obviously related to the unseen PSR. The double-cone
structure strongly suggests a two-sided collimated outflow from the
central PSR along a precessing beam. The diffuse emission shows a very
regular radial magnetic field structure as is clearly visible in the
Effelsberg 32~GHz image of polarized intensity. Fitting the radial
emissivity of G21.5$-$0.9  F\"urst et al. (\cite{fue88}) found as the
best solution a constant electron density with a radial linear decrease
of the magnetic field, although it must be noted that the fit is not
perfect. In any case, G21.5$-$0.9 defies the explanation in terms of an
expanding synchrotron nebula. Recent X-ray observations reveal weak
non-thermal diffuse emission extending fairly symmetric beyond the
previously known boundaries to about $150\arcsec$ (Slane et al.\
\cite{sla00}, Warwick et al.\ \cite{war01}). The emission is associated
with spectral softening of the emission.

G21.5$-$0.9 was detected in the infrared by Gallant \& Tuffs
(\cite{gal98}). They tried to describe available radio, infrared
and X-ray data with a broken power-law spectrum with a steeping from
$\alpha=0$ to $\alpha=-1$, where the break frequency is at about
200~GHz to 300~GHz. New radio observations with the Effelsberg,
Nobeyama and HHT telescopes between 43~GHz and 345~GHz indicate a break
frequency at about 20~GHz and a steepening by about $\Delta\alpha =
-0.36$ (Reich et al., in prep.), thus incompatible with a single
spectral break when including infrared as well as X-ray emission.
Since synchrotron aging cannot account for such a spectrum, intrinsic
properties of the electron spectrum or an unsteady evolution needs to
be considered.

Another plerion known since long to be different from Tau~A is 3C58.
Its diffuse emission with little small-scale structure is elliptical in
shape and shows almost no spatial variations in its spectrum. 3C58
originates from the A.D.~1181 SN event and is located at about 3.2~kpc
distance. Recently Camilo et al. (\cite{cam}) detected a very weak
65~ms radio PSR at the center of the SNR. The flux density of 3C58 was
reported to increase by about 0.3\%/year (Aller \& Reynolds\
\cite{all85}, Aller et al.\ \cite{all86}, Green\ \cite{gr87}),
which is difficult to understand within current models. 3C58 shows a
break in its spectrum around 50~GHz (Salter et al.\ \cite{sal89}),
which is confirmed by infrared data (Green \& Scheuer\ \cite{gre92}). An
image of the distribution of polarized intensity is shown in Fig.~15,
where despite of its rather smooth total-intensity distribution, very
patchy polarization is visible. Compared to G21.5$-$0.9 this reflects a
large difference in its internal magnetic field structure from an
undisturbed radial magnetic field. The 32~GHz image is rather similar
to the 1.4~GHz image of polarized intensities by Wilson \& Weiler
(\cite{wil76}) at about the same angular resolution. This indicates
little internal depolarization and the observed polarization structure
reflects regular magnetic field cells of the order of 10\% of the size
of the SNR. However, this does not imply a strongly enhanced magnetic
field in these regions because enhanced total intensities are not
observed.

\section{More radio observations}

High-frequency, high-resolution observations with polarization are of
particular importance for the identification of combined-type
and plerionic SNRs. It should be stressed again that despite all
efforts the number of known sources is quite small. High-sensitivity
observations are  required for the fainter sources of this class in
order to constrain their parameters in a comparable way. Faint
flat-spectrum cores seen in projection against steep-spectrum shell
emission is difficult to extract. The decomposition of diffuse and
small-scale emission provides hints towards the particle injection process and
the internal magnetic field structure. Details of the spectral
break, its relation to the different components and their spatial
variation are of interest to model the source evolution or to
constrain the particle injection spectrum including its time
dependence. Together with X-ray data from CHANDRA and XMM models of
the observational properties of this class of SNRs will be
substantially improved.

Despite the large number of known shell-type SNRs, there are still
ongoing identifications, in particular for large-diameter low-surface
brightness objects. To make such identifications at high frequencies
is difficult and rather time-consuming. A combination of
synthesis-telescope data with single-dish observations to reduce the
confusion limitations from extragalactic background sources is one way
to be successful at low frequencies.

\vskip 0.4cm

\begin{acknowledgements}
I like to thank Roland Kothes, B\"ulent Uyan{\i}ker and in particular
Ernst F\"urst for many years of common SNR research. Patricia Reich
and Ernst F\"urst critically read the manuscript. The support of 
SNR observations at the
Effelsberg 100-m telescope by Richard Wielebinski is acknowledged.
\end{acknowledgements}

\clearpage

\end{document}